\documentclass[preprint,review,12pt]{elsarticle}

\usepackage{graphics}
\usepackage{graphicx}
\usepackage{epsfig}

\usepackage{amssymb}
\usepackage{amsthm}

\usepackage{lineno}

\usepackage{color}

\journal{Annals of Physics}

\begin{document}

\begin{frontmatter}



\title{
von K\'arm\'an--Howarth and Corrsin equations closure based on Lagrangian  description of the fluid motion. 
}


\author{Nicola de Divitiis}
 
\address{"La Sapienza" University, Dipartimento di Ingegneria Meccanica e 
Aerospaziale, Via Eudossiana, 18, 00184 Rome, Italy, \\
phone: +39--0644585268, \ \ fax: +39--0644585750, \\ 
e-mail: n.dedivitiis@gmail.com, \  dedivitiis@dima.uniroma1.it,  \ nicola.dedivitiis@uniroma1.it}

\begin{abstract}
A new approach to obtain the closure formulas for the von K\'arm\'an--Howarth and Corrsin equations is presented, which is based on the Lagrangian representation of the fluid motion, and on the Liouville theorem associated to the kinematics of a pair of fluid particles.
This kinematics is characterized by the finite--scale separation vector which is assumed to be statistically independent from the velocity field. Such assumption is justified by the hypothesis of fully developed turbulence and by the property that this vector varies much more rapidly than the velocity field.
This formulation leads to the closure formulas of von K\'arm\'an--Howarth and Corrsin equations in terms of longitudinal velocity and temperature correlations following a demonstration completely different 
with respect to the previous works. 
Some of the properties and the limitations of the closed equations are discussed.
In particular, we show that the times of evolution of the developed kinetic energy and temperature spectra are finite quantities which depend on the initial conditions.
\end{abstract}

\begin{keyword}
von K\'arm\'an--Howarth equation, 
Corrsin equation, 
Liouville theorem,
Fully developed chaos,
Lyapunov exponent
\end{keyword}

\end{frontmatter}

\newcommand{\no}{\noindent}
\newcommand{\be}{\begin{equation}}
\newcommand{\ee}{\end{equation}}
\newcommand{\bea}{\begin{eqnarray}}
\newcommand{\eea}{\end{eqnarray}}
\newcommand{\bc}{\begin{center}}
\newcommand{\ec}{\end{center}}

\newcommand{\calr}{{\cal R}}
\newcommand{\calv}{{\cal V}}

\newcommand{\bff}{\mbox{\boldmath $f$}}
\newcommand{\bfg}{\mbox{\boldmath $g$}}
\newcommand{\bfh}{\mbox{\boldmath $h$}}
\newcommand{\bfi}{\mbox{\boldmath $i$}}
\newcommand{\bfm}{\mbox{\boldmath $m$}}
\newcommand{\bfp}{\mbox{\boldmath $p$}}
\newcommand{\bfr}{\mbox{\boldmath $r$}}
\newcommand{\bfu}{\mbox{\boldmath $u$}}
\newcommand{\bfv}{\mbox{\boldmath $v$}}
\newcommand{\bfx}{\mbox{\boldmath $x$}}
\newcommand{\bfy}{\mbox{\boldmath $y$}}
\newcommand{\bfw}{\mbox{\boldmath $w$}}
\newcommand{\bfk}{\mbox{\boldmath $\kappa$}}

\newcommand{\bfA}{\mbox{\boldmath $A$}}
\newcommand{\bfD}{\mbox{\boldmath $D$}}
\newcommand{\bfI}{\mbox{\boldmath $I$}}
\newcommand{\bfL}{\mbox{\boldmath $L$}}
\newcommand{\bfM}{\mbox{\boldmath $M$}}
\newcommand{\bfS}{\mbox{\boldmath $S$}}
\newcommand{\bfT}{\mbox{\boldmath $T$}}
\newcommand{\bfU}{\mbox{\boldmath $U$}}
\newcommand{\bfX}{\mbox{\boldmath $X$}}
\newcommand{\bfY}{\mbox{\boldmath $Y$}}
\newcommand{\bfK}{\mbox{\boldmath $K$}}
\newcommand{\bfR}{\mbox{\boldmath $R$}}

\newcommand{\bfrho}{\mbox{\boldmath $\varrho$}}
\newcommand{\bfchi}{\mbox{\boldmath $\chi$}}
\newcommand{\bfphi}{\mbox{\boldmath $\phi$}}
\newcommand{\bfPhi}{\mbox{\boldmath $\Phi$}}
\newcommand{\bflambda}{\mbox{\boldmath $\lambda$}}
\newcommand{\bfell}{\mbox{\boldmath $\ell$}}
\newcommand{\bfxi}{\mbox{\boldmath $\xi$}}
\newcommand{\bfeta}{\mbox{\boldmath $\eta$}}
\newcommand{\bfLambda}{\mbox{\boldmath $\Lambda$}}
\newcommand{\bfPsi}{\mbox{\boldmath $\Psi$}}
\newcommand{\bfXi}{\mbox{\boldmath $\Xi$}}
\newcommand{\bfomega}{\mbox{\boldmath $\omega$}}
\newcommand{\bfOmega}{\mbox{\boldmath $\Omega$}}
\newcommand{\bfeps}{\mbox{\boldmath $\varepsilon$}}
\newcommand{\bfepsn}{\mbox{\boldmath $\epsilon$}}
\newcommand{\bftau}{\mbox{\boldmath $\tau$}}
\newcommand{\bfzeta}{\mbox{\boldmath $\zeta$}}
\newcommand{\bfkappa}{\mbox{\boldmath $\kappa$}}
\newcommand{\bfsigma}{\mbox{\boldmath $\sigma$}}
\newcommand{\bftheta}{\mbox{\boldmath  $\vartheta$}}
\newcommand{\itPsi}{\mbox{\it $\Psi$}}
\newcommand{\itPhi}{\mbox{\it $\Phi$}}

\newcommand{\bint}{\mbox{ \int{a}{b}} }
\newcommand{\ds}{\displaystyle}
\newcommand{\Sum}{\Large \sum}



\bigskip

\section{Introduction}

Recently, a work dealing with the bifurcations analysis of the turbulent energy cascade \cite{deDivitiis_3} presents, among the other things, the relationship between Navier--Stokes equations bifurcations and turbulent energy cascade, showing that these bifurcations produce  
a negative value of the skewness of the longitudinal velocity difference, and a separation rate between contiguous trajectories which exponentially diverges with the time.

The present work proposes a specific analysis of the isotropic homogeneous turbulence of incompressible fluids based on Ref. \cite{deDivitiis_3}, and on the property that the contiguous fluid particles trajectories continuously diverge due to the Navier--Stokes bifurcations.
The proposed formulation adopts the Lagrangian representation of the fluid motion and the Liouville theorem.

According to the present study, the bifurcations determine the energy cascade, where the separation vector $\bfxi$ between two fluid particles trajectories varies much more rapidly than the velocity and temperature \cite{deDivitiis_3}. Such property, in conjunction with the hypothesis of fully developed turbulence, justifies the assumption that $\bfxi$ and the velocity field are statistically 
independent. This is the crucial hypothesis of the present analysis that allows, through the Liouville theorem, to analytically express the closure formulas of the von K\'arm\'an--Howarth and Corrsin equations in terms of longitudinal velocity and temperature correlations.
The proposed analysis also quantifies the mean effect of the trajectories separation through the average finite--scale Lyapunov exponent, a quantity depending on $r$ which gives the rate of separation of the trajectories with finite distance. This exponent is here calculated in function of the maximum finite--scale Lyapunov exponent through the distribution function of $\bfxi$ and the Liouville theorem.

{
For sake of the reader convenience, we report the main keypoints of the article: 
The first part of the work, devoted to the Lagrangian description of the fluid motion, 
provides the representation of velocity and temperature fields, and of
the kinematics of a pair of fluid particles. The Liouville theorem
 is  then introduced, and the statistical independence of $\bfxi$ from velocity and temperature
fields is properly justified.
In the second part, a relationship between average and maximal Lyapunov exponents of finite--scale, useful for the subsequent analysis, is determined, and the average Lyapunov exponent is expressed in terms of longitudinal velocity correlation.
Thereafter, the closure equations are obtained through the analytical elements introduced in the previous two parts.
}

The obtained results agree with those presented in Refs. \cite{deDivitiis_1, deDivitiis_2}, where the closure formulas are carried out using a fully different formulation and exploiting some of the properties of motion of the finite--scale Lyapunov basis, and the frame invariance 
of the triple correlations appearing in the von K\'arm\'an--Howarth and Corrsin equations.
This corroborates the results obtained in Refs. \cite{deDivitiis_1, deDivitiis_2}, showing the equivalence between the present approach and the analysis of Refs. \cite{deDivitiis_1, deDivitiis_2}.

Finally, some of the properties of the closed von K\'arm\'an--Howarth and Corrsin equations are studied and their limits of validity are discussed. Specifically, we show that the times of developing of velocity and temperature correlations are both finite quantities which depend on the initial conditions.

\bigskip

\section{ Background: longitudinal velocity and temperature correlations}

In fully developed homogeneous isotropic turbulence, the turbulent fluctuations of velocity and temperature are represented by the following pair correlation functions \cite{Karman38,  Corrsin_1}
\bea
\begin{array}{l@{\hspace{-0.cm}}l}
\ds f = 
\frac{\left\langle  u_r u_r' \right\rangle }{u^2}, \\\\
\ds f_\theta =
  \frac{\left\langle \vartheta \vartheta'\right\rangle}{\theta^2},
\end{array}
\label{ff}
\eea
where the brackets $\left\langle  \right\rangle$ denote the average calculated over the ensemble of velocity and temperature in the points $\bf x$ and ${\bf x} + \bf r$, 
$u_r = {\bf u}({\bf x}) \cdot {\bf r}/r$ and $u_r' = {\bf u}({\bf x}+ {\bf r}) \cdot {\bf r}/r$ are the longitudinal components of fluid velocities, { $\vartheta$ and $\vartheta'$ are the fluid temperatures in $\bf x$ and $\bf x + \bf r$},  and $u^2 = \langle {\bf u} \cdot {\bf u} \rangle/3$, $\theta^2 = \langle  \vartheta^2 \rangle$ represent the standard deviations of velocity and temperature, both constants in space due to homogeneity \cite{Karman38,  Corrsin_1}. 
Such correlations vary according to the von K\'arm\'an--Howarth and Corrsin equations \cite{Karman38, Batchelor53, Corrsin_1, Corrsin_2} which are the evolution equations for $f$ and $f_\theta$, respectively.
These equations, obtained through the momentum Navier--Stokes and the thermal energy equations,  are \cite{Karman38, Corrsin_1}
\bea
\ds \frac{\partial f}{\partial t} = 
\ds  \frac{K(r)}{u^2} +
\ds 2 \nu  \left(  \frac{\partial^2 f} {\partial r^2} +
\ds \frac{4}{r} \frac{\partial f}{\partial r}  \right) +\frac{10 \nu}{\lambda_T^2} f,
\label{karman-howarth}
\eea
\bea
\ds  \frac{\partial  f_\theta}{\partial t}  =
\ds  \frac{G(r)}{\theta^2} 
+2 \chi  \left( \frac{\partial^2 f_\theta}{\partial r^2}  
+ \frac{2}{r} \frac{\partial f_\theta}{\partial r} \right) 
 +  \frac{12 \chi}{\lambda_\theta^2} f_\theta, 
\label{corrsin}
\eea
where $\nu$ and $\chi = k/(\rho C_p)$ are kinematic viscosity and thermal diffusivity, $C_p$ and $k$ are specific heat at constant pressure and thermal conductivity, respectively.
The quantities $\lambda_T \equiv \sqrt{-1/f''(0)}$ and $\lambda_\theta \equiv \sqrt{-2/f_\theta''(0)}$ are the Taylor and Corrsin microscales.
The boundary conditions associated to Eqs. (\ref{karman-howarth}) and (\ref{corrsin}) are
\bea
\begin{array}{l@{\hspace{-0.cm}}l}
\ds f(0) = 1, \ \ \ \lim_{r \rightarrow \infty} f(r) =0, \\\\
\ds f_\theta(0) = 1, \ \ \ \lim_{r \rightarrow \infty} f_\theta(r) =0,
\end{array}
\label{bc}
\eea
In Eqs. (\ref{karman-howarth}) and (\ref{corrsin}), $K$ and $G$ arise from the inertia forces and from the convective terms of the Navier--Stokes and thermal energy equations \cite{Karman38,  Corrsin_1}, and can be expressed as
\bea
\begin{array}{l@{\hspace{-0.cm}}l}
\ds K = - 
\frac{\partial}{\partial r_k} \left\langle u_r u_r' \left( u_k'-u_k \right)  \right\rangle, \\\\
\ds G = - 
\frac{\partial}{\partial r_k} \left\langle \vartheta \vartheta' \left( u_k'-u_k \right)  \right\rangle,
\end{array}
\label{KG}
\eea
Accordingly, $K$ and $G$ do not modify neither the kinetic energy nor the thermal energy, and provide the energy cascade mechanism the effect of which vanishes for $r=0$ and for $r\rightarrow\infty$, i.e.
\bea
\begin{array}{l@{\hspace{-0.cm}}l}
\ds K(0) = 0, \ \ \lim_{r \rightarrow \infty} K(r) =0, \\\\
\ds G(0) = 0, \ \ \lim_{r \rightarrow \infty} G(r) =0
\end{array}
\label{KG0}
\eea
In this study, we analyze the case where $\nu$ does not depend on $\vartheta$, therefore
Eq. (\ref{karman-howarth}) is independent from Eq. (\ref{corrsin}) and $f_\theta$.
Conversely, the temperature fluctuations will depend on $\bf u$.

\bigskip

\section{Lagrangian representation of fluid motion, Liouville theorem \label{7} }

Following the present analysis, the turbulence is caused by the bifurcations of the Navier--Stokes equations, whereas the temperature plays the role of the passive scalar. This study applies also to any passive scalar which exhibits diffusivity.
These bifurcations frequentely occur in developed turbulence determining a 
condition of fully developed chaos where velocity and temperature exhibit chaotic fluctuations, and
the contiguous fluid particles trajectories diverge continuously with exponential growth rate  \cite{deDivitiis_3}. This implies that the fluid deformations are represented by exponential growth functions of the time, whereas the fluid state variables, like velocity and temperature, are slow growth functions of $t$.

\bigskip

\subsection{Representation of velocity and temperature fields}

To represent velocity and temperature fluctuations, we start from Ref. \cite{deDivitiis_3}, where the Navier--Stokes equations are written in the symbolic form of operators. 
Here, the Navier--Stokes equations are expressed following the Lagrangian description of motion 
\cite{Truesdell77}, and to analyze the temperature fluctuations, the thermal energy equation is also given in the same form
\bea
\begin{array}{l@{\hspace{-0.cm}}l}
\ds \dot{\bf u} = {\bf N}({\bf u}) \equiv 
{\bf N}_p({\bf u}) +\nu {\bf L_N} {\bf u} , \\\\
\ds \dot{\bftheta} = {\bf M}({\bftheta}) \equiv 
 \chi {\bf L_M} {\bftheta} 
\end{array}
\label{NS_Tee}
\eea
The first equation is the Navier--Stokes equations in the symbolic form of operators \cite{Tsinober2009}, where the pressure field $p$ has been eliminated through the continuity equation, 
$\bf u$ and $\dot{\bf u}$ denote, respectively, velocity and acceleration fields \cite{Truesdell77}, ${\bf L_N} (\circ) \equiv$ ${\bf L_M}(\circ) \equiv \nabla^2(\circ)$, and
${\bf N}_p({\bf u}) \equiv - \nabla p/\rho$. 
Specifically, ${\bf N}_p({\bf u})$ is the integral non--linear operator which expresses the pressure gradient through the velocity field in the entire fluid domain.
That is, the pressure gives the non--local effect of the velocity field 
\cite{Tsinober2009}, and the Navier--Stokes equations are reduced to be an integro--differential equation formally expressed by Eq. (\ref{NS_Tee}) in the symbolic form of operators. 

{ The other equation describes the fluid temperature variations and is also in the symbolic form of operators. The temperature field $\bftheta$ is denoted in bold type as it is an element of the vector space of the temperature fields $\left\lbrace \bftheta \right\rbrace$ which is an infinite dimensional manifold, thus $\vartheta$ and $\vartheta'$ are the values of $\bftheta$ calculated in $\bf x$ and $\bf x + \bf r$, respectively, and $\dot{\bftheta}$ is the material time derivative of temperature \cite{Truesdell77}}.

\bigskip

Now, the bifurcations are responsible for the fluctuations of $\bf u$ and $\bftheta$ whose
statistics is formally described by the distribution $F$=$F\left\lbrace {\bf u}, {\bftheta} \right\rbrace$, a functional of $\bf u$ and $\bftheta$ which satisfies the Liouville theorem associated to Eqs. (\ref{NS_Tee})
\bea
\begin{array}{l@{\hspace{-0.cm}}l}
\ds \frac{\partial F}{\partial t}+
\nabla_{\bf u} \cdot \left( {\bf N} F \right)+
\nabla_{\bf \vartheta} \cdot \left( {\bf M} F \right)=0
\end{array}
\label{Liouville NSE}
\eea
{ where the operators $\nabla_{\bf u} \cdot(\circ)$ and $\nabla_{\vartheta} \cdot(\circ)$ denote the divergence in the vector spaces of velocity and temperature fields}.
This theorem is derived from Eq. (\ref{NS_Tee}) and from the condition that the integral of $F$ over $\cal V$ identically equals the unity
\bea
\begin{array}{l@{\hspace{-0.cm}}l}
\ds \int_{\cal V}  F \ d {\cal V} =1, \ \ \forall t \ge 0.
\end{array}
\label{c0}
\eea
where ${\cal V} = \left\lbrace {\bf u} \right\rbrace \times \left\lbrace {\bftheta} \right\rbrace$
is the phase space { associated to Eq. (\ref{NS_Tee})}, in which $\left\lbrace {\bf u} \right\rbrace$ and  $\left\lbrace {\bftheta} \right\rbrace$ are the { vector spaces} of velocity and temperature fields, and $d {\cal V} = d {\bf u} \ d {\bftheta}$ { represents the elemental volume in the phase space ${\cal V}$.}

\bigskip

\subsection{Representation of fluid kinematics. Kinematics of a pair of fluid particles}

To complete the Lagrangian description of fluid motion, the kinematics equations are added to 
Eqs. (\ref{NS_Tee}).
To our purposes, we consider the trajectories of fluid particles passing through $\bf x$ and 
${\bf x} +{\bf r}$ at $t=$0, whose evolution equations are
\bea
\begin{array}{l@{\hspace{-0.cm}}l}
\dot{\bfx} = {\bf u} ({\bfx}, t), \\\\
\dot{\bfxi} = {\bf u} ({\bfx}+{\bfxi}, t)- {\bf u} ({\bfx}, t),   \\\\
\end{array}
\label{lb}
\eea
with the initial conditions
\bea
\begin{array}{l@{\hspace{-0.cm}}l}
{\bfx}(0)  = {\bf x}, \\\\
{\bfxi}(0)  = {\bf r}
\end{array}
\label{ic}
\eea
where $\bfxi$ is the separation vector {and ${\bfxi}(0)  = {\bf r}$ is its initial value}. 
This separation vector is a function of $t$ of exponential growth \cite{deDivitiis_3} which depends on the scale $r = \vert {\bf r} \vert$ and on the initial condition ${\bfxi}(0)$, whereas $\bf u$ varies according to Eq. (\ref{NS_Tee}). 
Following Eq. (\ref{lb}), the map { ${\bfchi}: {\bfxi(0)} \rightarrow {\bfxi}(t)$} provides the current separation distance $\bfxi$ between two fluid particles which are located at the referential positions $\bf x$ and $\bf x + r$ at t = 0 \cite{Truesdell77}.
{It is worth to remark that $\bf x$ and $\bf x + r$ are fixed points of the inertial frame, whereas $\bfx$ and $\bfx + \bfxi$ represent the positions of the fluid particles which vary with the time and that satisfy the initial conditions (\ref{ic}).}

To analyze the average effect of the  trajectories separation, the statistics of $\bfxi$ is now introduced.
This statistics is described by the distribution function of $\bfxi$ which does not depend on $\bfx$ 
due to turbulence homogeneity
\bea
\begin{array}{l@{\hspace{-0.cm}}l}
\ds P=P(t, {\bfxi})
\end{array}
\eea
where
\bea
\begin{array}{l@{\hspace{-0.cm}}l}
\ds  \int_{\bfXi} P \  d {\bfXi} = 1, \ \ \forall t \geq 0
\end{array}
\label{intP=1}
\eea
in which ${\bfXi}$ $=\left\lbrace {\bfx} \right\rbrace$ $\times \left\lbrace {\bfxi} \right\rbrace$,
and $d{\bfXi} = d {\bfx} \ d {\bfxi}$ represents the relative elemental volume.
As the problem (\ref{lb}) is studied in an infinite fluid domain, $P$ identically vanishes on the boundaries of  $\bfXi$, i.e. 
\bea
\begin{array}{l@{\hspace{-0.cm}}l}
\ds P(t, {\bfxi}) = 0 \ \ \ \forall  \left( {\bfx}, {\bfxi}\right)  \in \partial 
 {\bfXi}, \ \ \forall t \geq 0
\end{array}
\label{Pborder}
\eea
Next, this PDF satisfies the Liouville theorem associated to Eqs. (\ref{lb}) and (\ref{intP=1})
\bea
\begin{array}{l@{\hspace{-0.cm}}l}
\ds \frac{\partial P}{\partial t} 
\ds + \frac{\partial}{\partial {\bfxi}} \cdot \left( P \dot{\bfxi} \right) 
\ds + \frac{\partial}{\partial {\bfx}} \cdot \left( P \dot{\bfx} \right) 
=0
\end{array}
\label{Liouville0}
\eea
where the differential operators
${\partial}/{\partial {\bfxi}} \cdot (\circ)$ and ${\partial}/{\partial {\bfx}} \cdot (\circ)$ indicate the divergence of $(\circ)$ with  respect to $\bfxi$ and $\bfx$, respectively.
Taking into account the hypothesis of homogeneity and the continuity equation, the Liouville equation
reads as
\bea
\begin{array}{l@{\hspace{-0.cm}}l}
\ds \frac{\partial P}{\partial t} 
\ds + \frac{\partial}{\partial {\bfxi}} \cdot \left( P \dot{\bfxi} \right) 
=0
\end{array}
\label{Liouville}
\eea
As the initial condition of $\bfxi$ and $\bfx$ is given by Eq. (\ref{ic}), the initial condition of  the distribution function $P=P(t, {\bfxi})$ is 
\bea
\ds P(0, {\bfxi})= \delta({\bfxi}, {\bf r})
\label{ic_P}
\eea
where $\delta(.,.)$ is the Dirac's delta.

Therefore, the average of any quantity depending on $\bfx$ and $\bfxi$, 
say $\psi = \psi({\bfx}, {\bfxi})$, will be calculated as integral of $P$ over $\left\lbrace \bfXi \right\rbrace$, according to
\bea
\begin{array}{l@{\hspace{-0.cm}}l}
\ds \overline{\psi} = \int_{\bfXi} P \ \psi \  d {\bfXi}
\end{array}
\label{average}
\eea

\bigskip

\subsection{Statistical independence of $\bfxi$ from $\bf u$ and $\bftheta$}

At this stage of the analysis, it is worth to remark that $\bfxi$, { directly responsible for the relative trajectories mixing \cite{Ottino90},} behaves like a function of fast growth of $t$, therefore it varies much more rapidly than $\bf u$ and $\bftheta$ \cite{deDivitiis_3}.
{ Accordingly, in fully developed turbulence, the time--scales of $\bfxi$ are expected to be completely separated from those associated to $\bf u$ and $\bftheta$ in the sense that 
$\bfxi$ and ($\bf u$, $\bftheta$) exhibit chaotic behavior and their power spectra 
are supposed to be located in frequency intervals which are completely separated. 
For this reason, $\bfxi$ and ($\bf u$, $\bftheta$) are considered to be statistically uncorrelated.
From the physical point of view, this means that the effect of the trajectories mixing is much more rapid and statistically uncorrelated with respect to the dynamics of the fluid system.
This property is supported by the arguments presented in Ref. \cite{Ottino90} (and references therein), where the author observes the that:
a) The fields ${\bf u} ({\bfx}, t)$, (and therefore also ${\bf u} ({\bfx+ \bfxi}, t)- {\bf u} ({\bfx}, t)$) produce chaotic trajectories also for relatively simple mathematical structure of the right--hand sides ${\bf u} ({\bfx}, t)$ (also for steady fields!). 
b) The flows given by ${\bf u} ({\bfx}, t)$ (and therefore also ${\bf u} ({\bfx+ \bfxi}, t)- {\bf u} ({\bfx}, t)$) stretch and fold continuously and rapidly causing an effective mixing of the particles trajectories.
}
Hence, it is reasonable to postulate  that $\bfxi$ and ($\bf u$, $\bftheta$) are statistically uncorrelated, and that all the quantities depending on $\bfxi$ are fast variables whose average is calculated through Eq. (\ref{average}).
This is the fundamental hypothesis of the present work following  which the distribution function of $\bfxi$, $\bf u$ and $\bftheta$ is the product of $P$ and $F$: 
\bea
{\cal F} (t, {\bfxi}, {\bf u}, {\bftheta}) = 
P(t, {\bfxi}) \
F(t, {\bf u}, {\bftheta})
\label{stat_indep}
\eea 
The time derivative of ${\cal F}$ is then
\bea
\begin{array}{l@{\hspace{-0.cm}}l}
\ds \frac{\partial {\cal F}}{\partial t} =
\ds \frac{\partial {F}}{\partial t} \ P +
\ds \frac{\partial {P}}{\partial t} \ F 
\end{array}
\label{df_dt}
\eea
The first term at the R.H.S. of Eq. (\ref{df_dt}) is due to the time variations of $\bf u$ and $\bftheta$ following Eqs. (\ref{NS_Tee}), thus such term gives the time variations of kinetic and thermal energies. As far as the second term is concerned, it is related only to the fluid kinematics in line with Eq. (\ref{Liouville}), thus it does not contribute to kinetic and thermal energies variations.

{ We conclude this section by observing the limits of validity of Eq. (\ref{stat_indep}). As these limits arise from the hypothesis of fully developed turbulence, in the cases of intermediate stages of turbulence or in decaying turbulence, the statistical independence of $\bfxi$ and $\bf u$ could be not verified.}

\bigskip

\section{Average finite--scale Lyapunov exponent in terms of the maximal finite--scale Lyapunov exponent}

{ The purpose of this section is to present the link between average and maximum
Lyapunov exponents of finite--scales that will be useful later on in this analysis.}

The average effect of the trajectories divergence is quantified by the 
average finite--scale Lyapunov exponent, a quantity which expresses the separation rate between two trajectories with finite distance.
To define this exponent, observe that the following quantity
\bea
\ds {\lambda}'= \frac{1}{2} \frac{d}{dt} {\ln({\bfxi}\cdot{\bfxi})}
\eea
gives the local growth rate of  $\vert \bfxi \vert$.
Following its definition, the maximal finite--scale Lyapunov exponent gives the maximum growth rate of the trajectories distance. In the present framework, this exponent can be written as 
\bea
\ds \lambda= \frac{1}{2} \frac{d}{dt} \overline{\ln({\bfxi}\cdot{\bfxi})}   
\label{max lyap exp}
\eea
The overline $\ds \overline{\circ}$ denotes the average calculated according to 
Eq. (\ref{average}), where $\lambda=\lambda(r)$ thanks to the turbulence isotropy. 
This exponent is associated to a given spatial direction $l_1$, variable with the time,
which corresponds to the maximum growth rate of trajectories distance.
\begin{figure}[h]
\centering
e\includegraphics[width=0.40\textwidth]{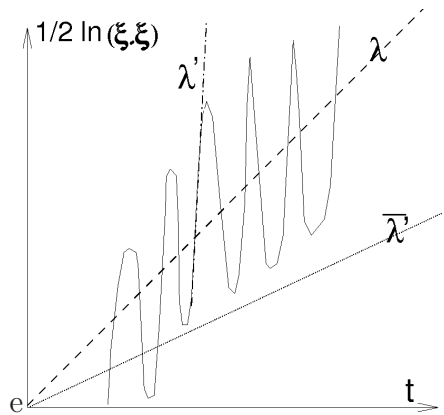}
\caption{Scheme of definition of $\lambda'$, $\lambda$ and $\overline{\lambda'}$.}
\label{figura_0}
\end{figure}

On the other hand, the trajectories separation happens with a mean growth rate. This 
latter is defined through the average finite--scale Lyapunov exponent, a quantity calculated as the mean of $\lambda'$
\bea
\ds \overline{\lambda'} = \lim_{T \rightarrow \infty} 
\frac{1}{T} \int_0^T 
\frac{\dot{\bfxi}\cdot{\bfxi}}{{\bfxi}\cdot{\bfxi}}  \ dt
\label{average lambda 0}
\eea
If the ergodic hypothesis is satisfied, this average exponent is calculated in terms of the distribution $P$ as
\bea
\begin{array}{l@{\hspace{-0.cm}}l}
\ds \overline{\lambda'} 
\equiv 
\int_{\bfXi} P \ \frac{\dot{\bfxi}\cdot{\bfxi}}{{\bfxi}\cdot{\bfxi}} 
\ d {\bfXi} = 
 \frac{1}{2} \int_{\bfXi} P \frac{d}{dt} \left(  \ln({\bfxi}\cdot{\bfxi}) \right) 
\ d {\bfXi} 
\end{array}
\label{average lambda}
\eea
With reference to Fig. \ref{figura_0}, the continuous line represents the time variations of
$1/2 {\ln({\bfxi}\cdot{\bfxi})}$ and its local slope gives $\lambda'$ (dash--dotted line), whereas $\lambda$ and $\overline{\lambda'}$ are represented by dashed and dotted lines, respectively.

The two exponents are related with each other, and to express $\overline{\lambda'}$ in function of ${\lambda}$, we write $\overline{\lambda'}$ in terms of $\partial P/\partial t$ 
\bea
\begin{array}{l@{\hspace{-0.cm}}l}
\ds \overline{\lambda'} = \frac{1}{2} \frac{d}{dt} \overline{\ln({\bfxi}\cdot{\bfxi})} - \frac{1}{2} \int_{\bfXi}  \frac{\partial P}{\partial t}       \ln({\bfxi}\cdot{\bfxi})
\ d {\bfXi}
\end{array}
\label{appo1}
\eea
where the first term at the R.H.S. of Eq. (\ref{appo1}) identifies $\lambda$, 
whereas the second one is expressed by means of the Liouville theorem
\bea
\begin{array}{l@{\hspace{-0.cm}}l}
\ds \overline{\lambda'} = \lambda + \frac{1}{2} \int_{\bfXi} 
\frac{\partial}{\partial {\bfxi}} \cdot \left( P \dot{\bfxi} \right)
 \ln({\bfxi}\cdot{\bfxi})
\ d {\bfXi}
\end{array}
\label{appo2}
\eea
Integrating by parts the last term at the R.H.S. of Eq. (\ref{appo2}), and using the Green's second identity, we obtain the sum of two terms. The first one of these identically vanishes as it is an integral of $P$ over $\partial \left\lbrace {\bfxi} \right\rbrace$ where $P\equiv$ 0, whereas the other one is different from zero. This leads to 
\bea
\begin{array}{l@{\hspace{-0.cm}}l}
\ds \overline{\lambda'} = \lambda -
\int_{\bfXi} P \ \frac{\dot{\bfxi}\cdot{\bfxi}}{{\bfxi}\cdot{\bfxi}} 
\ d {\bfXi}
\end{array}
\eea
Hence, taking into account Eq. (\ref{average lambda}), we have
\bea
\begin{array}{l@{\hspace{-0.cm}}l}
\ds \overline{\lambda'} = \frac{\lambda}{2} 
\end{array}
\label{rate}
\eea
Due to homogeneity, the separation rate does not depend on $\bf x$, and thanks to 
the isotropy $\overline{\lambda'}$ and ${\lambda}$ are both even functions of $r$ 
\cite{Batchelor53, Robertson40}.

Finally, it is worth remarking a property of the function ${\bfchi}:  {\bfxi}(0) \rightarrow {\bfxi}$: when $0 < t \lesssim  1/\lambda$, $\xi \rightarrow r \exp({\lambda} t)$, thus ${\bfchi}$ satisfies the following equation
\bea
\ds \frac{\partial \xi}{\partial r} = \frac{\xi}{r}, \ \ 0 < t \lesssim 1/\lambda
\label{xir}
\eea

\bigskip

\section{Average Lyapunov exponent in terms of velocity correlation }

In order to express $\overline{\lambda'}$ in terms of $f$, observe that along the direction $l_1$,
$\bfxi$ varies following $\xi=\xi(0) \exp(\lambda t)$, i.e.
\bea
{\bfxi} = {\bf Q}(t) {\bfxi}(0) \exp(\lambda t)
\eea
where ${\bf Q}(t)$ is a proper rotation matrix which gives the orientation of $\bfxi$ with respect to the inertial reference frame. Hence, the relative velocity $ \dot{\bfxi}$ is 
\bea
\ds \dot{\bfxi} = {\bfomega}_{\bflambda} \times {\bfxi} + \lambda {\bfxi}, 
\eea 
where ${\bfomega}_{\bflambda}$ is the angular velocity of $\bfxi$ and $l_1$ with respect to
the inertial reference.

Now, because of the isotropy, the standard deviation of $\Delta u_r$  does not depend on ${\bf r}/r$, and this implies that 
\bea
\ds \left\langle (u_r'-u_r)^2 \right\rangle =  
\left( \dot{\bfxi} \cdot \frac{{\bfxi}}{\xi} \right)_{t=0}^2  \equiv \lambda^2 r^2,
\label{isotropy}
\eea
Hence, $f$ and  $\overline{\lambda'}$  are linked with each other according to 
\bea
\ds \overline{\lambda'}(r) = \frac{u}{r} \sqrt{\frac{1-f}{2}}
\label{lambda}
\eea
Equation (\ref{lambda}) is defined only for $f<1$
and this agrees with Schwarz inequality \cite{Batchelor53}, and in particular, 
for $\partial f /\partial r <$0 near the origin. 

In conclusion, in fully developed homogeneous isotropic turbulence, the fluid particles trajectories continuously diverge with an average separation rate $\dot{R}$ which depends on $r$ following
\bea
\ds \dot{R} =\overline{\lambda'} r = u \sqrt{\frac{1-f}{2}}
\label{Rdot}
\eea

\bigskip

\section{Closure equations}

{ Here, the closure formulas of Eqs. (\ref{karman-howarth}) and (\ref{corrsin}) are
determined by means of the elements introduced in the previous sections.}

To obtain these formulas,
observe that $K$ and $G$ are responsible only for the energy cascade.
This latter is the kinetic and thermal energy flow between the length scales, without changing the total amount of kinetic and thermal energy (see for instance \cite{Batchelor53, Corrsin_1} and references therein). 
In fact, inertia and pressure forces and the convective term produce an interaction between
the various Fourier components of velocity and temperature spectra which gives the kinetic and thermal energy transfer between the volume elements in the wave--number space, where the global effect of such interaction leaves $u^2$ and $\theta^2$ unaltered \cite{Batchelor53, Corrsin_1}.

Now, to determine $K$ and $G$, we first calculate the pair correlations $f$ and $f_\theta$
through the PDF $\cal F$, taking into account the initial condition of $P$ given
by Eq. (\ref{ic_P}). Accordingly, longitudinal velocity and temperature correlations 
can be expressed through ${\cal F}$ 
\bea
\begin{array}{l@{\hspace{-0.cm}}l}
\ds u^2 f(r) = \left\langle u_r u_r' \right\rangle = \int_{\bfXi} \int_{\cal V} {\cal F}  \ 
u_\xi({\bfx}) u_\xi({\bfx}+{\bfxi}) \ d {\cal V} d {\bfXi} 
= \int_{\cal V} F \ u_r u_r'  \ d {\cal V}, \\\\
\ds \theta^2 f_\theta(r) = \left\langle \vartheta \vartheta' \right\rangle = \int_{\bfXi} \int_{\cal V} {\cal F}  \ \vartheta({\bfx}) \vartheta({\bfx}+{\bfxi}) \ d {\cal V} d {\bfXi}
= \int_{\cal V} F \ \vartheta \vartheta'  \ d {\cal V},
\end{array}
\label{fr}
\eea
or in terms of $F$ as
\bea
\begin{array}{l@{\hspace{-0.cm}}l}
\ds u^2 f(\xi) = \left\langle u_\xi({\bfx}) u_\xi({\bfx}+{\bfxi}) \right\rangle = \int_{\cal V} F  \ 
u_\xi({\bfx}) u_\xi({\bfx}+{\bfxi}) \ d {\cal V}, \\\\
\ds \theta^2 f_\theta(\xi) = \left\langle \vartheta({\bfx}) \vartheta({\bfx}+{\bfxi}) \right\rangle = \int_{\cal V} F  \ \vartheta({\bfx}) \vartheta({\bfx}+{\bfxi}) \ d {\cal V},
\end{array}
\label{fxi}
\eea
in which $u_\xi = {\bf u}\cdot {\bfxi}/\xi$.
The correlations (\ref{fr}) and (\ref{fxi}) are linked with each other because the fluctuation  ${\bfchi}:$ $ {\bfxi}(t^-)$ $\rightarrow$ $\bfxi(t)$, ${\bfxi}(t^-)={\bf r}$, $t=t^- +\varepsilon$,
$\varepsilon>0$ varying much more rapidly than $f$ and $f_\theta$, leaves unaltered the correlations in the transformed points, therefore, taking into account  
Eq. (\ref{xir}), we have $f(t, \xi)$=$f(t^-, r)$, $f_\theta(t, \xi)$=$f_\theta(t^-, r)$, and
\bea
\begin{array}{l@{\hspace{-0.cm}}l}
\ds  \left(  \frac{\partial f(\xi)}{\partial \xi} \right)_t
=  \left( \frac{\partial f(r)}{\partial r} \right)_{t^-} \ \frac{\partial r}{\partial \xi} 
= \left( \frac{\partial f(r)}{\partial r} \right)_{t^-} \ \frac{r}{\xi}, \\\\ 
\ds \left( \frac{\partial f_\theta(\xi)}{\partial \xi} \right)_t
=  \left( \frac{\partial f_\theta(r)}{\partial r}\right)_{t^-} \  \frac{\partial r}{\partial \xi} 
= \left( \frac{\partial f_\theta(r)}{\partial r}\right)_{t^-} \ \frac{r}{\xi}
\end{array}
\label{frxi}
\eea

Hence, the time derivatives of the correlation functions are calculated by means of 
 $\partial {\cal F}/\partial t$, using Eq. (\ref{df_dt}) 
\bea
\begin{array}{l@{\hspace{-0.cm}}l}
\ds \frac{\partial}{\partial t} \left\langle u_r u_r' \right\rangle =
\int_{\bfXi} \int_{\cal V} 
\left(  \frac{\partial {F}}{\partial t} \ P + \ds \frac{\partial {P}}{\partial t} \ F \right)  
\ u_\xi({\bfx}) u_\xi({\bfx}+{\bfxi}) \ d {\cal V} d {\bfXi} \\\\
\ds \frac{\partial}{\partial t} \left\langle \vartheta \vartheta' \right\rangle =
\ds \int_{\bfXi} \int_{\cal V} 
\left(  \frac{\partial {F}}{\partial t} \ P + \ds \frac{\partial {P}}{\partial t} \ F \right) \ 
 \vartheta({\bfx}) \vartheta({\bfx}+{\bfxi}) \ d {\cal V} d {\bfXi}
\end{array}
\eea
As previously seen, the terms with $\partial F/ \partial t$ correspond to the
time variations of kinetic and thermal energies and to the changing of the correlations caused by
viscosity and thermal conductivity, whereas those with $\partial P/ \partial t$, arising from the fluctuations of $\bfxi$, do not modify neither $u^2$ nor $\theta^2$.
This identifies $K$ and $G$ as 
\bea
\begin{array}{l@{\hspace{-0.cm}}l}
\ds K =
\int_{\bfXi} \int_{\cal V} 
\frac{\partial {P}}{\partial t} \ F \  u_\xi({\bfx}) u_\xi({\bfx}+{\bfxi})  \ d {\cal V} d {\bfXi} \\\\
\ds G  =
\int_{\bfXi} \int_{\cal V} 
\frac{\partial {P}}{\partial t} \ F \  \vartheta({\bfx}) \vartheta({\bfx}+{\bfxi}) \ d {\cal V} d {\bfXi}
\end{array}
\eea
where $\partial P/ \partial t$ is expressed through the Liouville theorem (\ref{Liouville}) 
\bea
\begin{array}{l@{\hspace{-0.cm}}l}
\ds K =
- \int_{\bfXi} \int_{\cal V} 
\frac{\partial}{\partial {\bfxi}} \cdot \left( P \dot{\bfxi} \right)  
\ F \ u_\xi({\bfx}) u_\xi({\bfx}+{\bfxi}) \ d {\cal V} d {\bfXi} \\\\
\ds G  =
- \int_{\bfXi} \int_{\cal V} 
\frac{\partial}{\partial {\bfxi}} \cdot \left( P \dot{\bfxi} \right)  
 \ F \ \vartheta({\bfx}) \vartheta({\bfx}+{\bfxi}) \ d {\cal V} d {\bfXi}
\end{array}
\label{appoII}
\eea
Next, integrating by parts the R.H.S. of Eq. (\ref{appoII}), and using the Green's second identity, we obtain that both $K$ and $G$ are the sum of two terms. The first ones identically vanish as these are integrals of $P$ over $\partial \left\lbrace {\bfxi}\right\rbrace $ in which $P\equiv$ 0, whereas the other ones are different from zero. 
This leads to 
\bea
\begin{array}{l@{\hspace{-0.cm}}l}
\ds K =
 \int_{\bfXi}  
P \ \dot{\bfxi} \cdot \frac{\partial}{\partial \bfxi} \left\langle  u_\xi({\bfx}) u_\xi({\bfx}+{\bfxi}) \right\rangle  \ d {\bfXi} \\\\
\ds G  =
 \int_{\bfXi}  
P \ \dot{\bfxi} \cdot  \frac{\partial}{\partial \bfxi} \left\langle \vartheta({\bfx}) \vartheta({\bfx}+{\bfxi}) \right\rangle \ d {\bfXi} 
\end{array}
\label{KG_0}
\eea
Taking into account the turbulence isotropy, and Eq. (\ref{frxi}), we obtain
\bea
\begin{array}{l@{\hspace{-0.cm}}l}
\ds  \frac{\partial} {\partial \bfxi} \left\langle  u_\xi({\bfx}) u_\xi({\bfx}+{\bfxi})\right\rangle =
 \frac{\partial}{\partial \xi} \left\langle 
u_\xi({\bfx}) u_\xi({\bfx}+{\bfxi}) \right\rangle  \frac{\bfxi}{\xi}=
\frac{\partial}{\partial r} \left\langle u_r u_r'\right\rangle \ \frac{{\bfxi}} {{\bfxi}\cdot{\bfxi}}  \ r, \\\\
\ds  \frac{\partial}{\partial \bfxi} \left\langle  \vartheta({\bfx}) \vartheta({\bfx}+{\bfxi})\right\rangle  =
 \frac{\partial}{\partial \xi} \left\langle \vartheta({\bfx}) \vartheta({\bfx}+{\bfxi})\right\rangle
\frac{\bfxi}{\xi}=
\frac{\partial}{\partial r} \left\langle \vartheta \vartheta'\right\rangle \ \frac{{\bfxi}}{{\bfxi}\cdot{\bfxi}} \ r
\end{array} 
\label{fluc}
\eea
It is worth remarking that $\dot{\bfxi}$, depending on $\bfx$ and $\bfxi$, is considered to be a fast variable, thus its contribution is calculated with the statistics of $P$. 
This leads to express $K$ and $G$ in terms of $\partial f / \partial r$ and $\partial f_\theta  / \partial r$
\bea
\begin{array}{l@{\hspace{-0.cm}}l}
\ds K  = u^2 \frac{\partial f} {\partial r} r \int_{\bfXi} P \ \frac{\dot{\bfxi}\cdot{\bfxi}}{{\bfxi}\cdot{\bfxi}} 
\ d {\bfXi}, \\\\
\ds G  =  \theta^2  \frac{\partial f_\theta} {\partial r} r \int_{\bfXi} P \ \frac{\dot{\bfxi}\cdot{\bfxi}}{{\bfxi}\cdot{\bfxi}} 
\ d {\bfXi}
\end{array}
\label{K0G00}
\eea 
Hence, in view of Eqs. (\ref{average lambda}) and (\ref{lambda}), we obtain the following expressions
\bea
\begin{array}{l@{\hspace{-0.cm}}l}
\ds K  = u^3 \sqrt{\frac{1-f}{2}} \ \frac{\partial f} {\partial r}, \\\\
\ds G  =  \theta^2 u \sqrt{\frac{1-f}{2}} \ \frac{\partial f_\theta} {\partial r}
\end{array}
\label{K0G0}
\eea
Following Eqs. (\ref{K0G0}), the phenomenon of kinetic and thermal energy cascade is due to the combined action of the spatial variations of $f$, $f_\theta$ and of the trajectories separation.
Therefore, $K$ and $G$ result to be non--monotonic even functions of $r$ which vanish for $r=0$ 
and tend to zero as $r \rightarrow \infty$. In particular, $K$ gives also the values of the skewness of $\partial u_r/\partial r$  \cite{deDivitiis_1} which is constant and is equal to $-3/7$.
{ Table 1 reports the comparison between the value of the skewness 
\bea
\ds H_3(0)= \frac{\langle(\partial u_r/\partial r)^3 \rangle}{\langle (\partial u_r/\partial r)^2\rangle^{3/2}}
\nonumber
\eea 
calculated with the proposed expression of $K$ and those obtained by the several authors
with direct numerical simulation of the Navier--Stokes equations (DNS), and Large--eddy simulations (LES). It results that the maximum absolute difference between the proposed value and the other ones results to be less than 10 $\%$. Other comparisons between the proposed closure formulas and the results of the literature can be found in the previous works \cite{deDivitiis_1, deDivitiis_2, deDivitiis_4}.}
{
\begin{table}[t]
\centering
\caption{Comparison of the results: Skewness of $\partial u_r/ \partial r$ at diverse Taylor--scale Reynolds number $R_T$ following different authors.}
\vspace{2. mm}
\begin{tabular}{cccc} 
\hline
\hline
Reference   &  Simulation  &  $R_T$  & $H_3(0)$  \\
\hline 
Present result      &  -  &  -    & -3/7 = -0.428...  \\
\cite{Orszag72}      & DNS &    45 & -0.47    \\
\cite{Panda89}      & DNS &    64 & -0.40    \\
\cite{Chen92}       & DNS &    202 & -0.44   \\ 
\cite{Carati95}     & LES &   $\infty$     &  -0.40 \\  
\cite{Anderson99}   & LES &    $<$ 71 &  -0.40          \\ 
\cite{Kang2003}     & LES &  720        &    -0.42      \\ 
\hline
\hline
 \end{tabular}
\label{table1}
\end{table} 
}

Equations (\ref{K0G0}) coincide with those obtained in Refs. \cite{deDivitiis_1, deDivitiis_2} 
through the properties of the motion of the finite--scale Lyapunov basis and of the frame invariance of $K$ and $G$. This shows that the hypotheses adopted in Refs. \cite{deDivitiis_1, deDivitiis_2} agree with the assumptions of the present analysis.

The main asset of Eqs. (\ref{K0G0}) with respect to the other models is that Eqs. (\ref{K0G0}) are not based on phenomenological assumptions, such as for instance, the existence of the eddy viscosity 
\cite{Hasselmann58, Millionshtchikov69, Oberlack93, Mellor84, Baev, Antonia2013}, 
but are obtained through theoretical considerations regarding the statistical independence of $\bfxi$ from $\bf u$, and the Liouville theorem. Thanks to their theoretical foundation, Eqs. (\ref{K0G0}) do not incorporate free model parameters or empirical constants which have to be identified.

Refs. \cite{deDivitiis_1, deDivitiis_2} show that such formulas describe adequately the energy cascade mechanisms. Specifically, $K$ reproduces the phenomenon of kinetic energy cascade in line with the Kolmogorov law, and $G$ describes the thermal energy cascade according to the theoretical argumentation presented in the Refs. \cite{Batchelor_2, Batchelor_3, Obukhov}, to experimental results \cite{Gibson, Mydlarski}, and to  numerical data \cite{Rogallo, Donzis}.

$K$ and $G$ arise from inertia forces and convective term, thus these do not depend on $\nu$ and $\chi$. Specifically, $K$ and $G$ are only indirectly related to $\nu$ and $\chi$  through $f$ and $f_\theta$ whose time evolutions depend on diffusivities  by means of 
Eqs. (\ref{karman-howarth}) and (\ref{corrsin}).

{ Finally, we remark the limits of the proposed equations. Such limits arise from the hypotheses under which Eqs. (\ref{K0G0}) are derived:
Eqs. (\ref{K0G0}) hold only in regime of fully developed turbulence where the flow statistical properties exhibit homogeneity and isotropy. Otherwise, during the transition through the 
intermediate stages of the turbulence, in decaying turbulence, or in more complex situations of developed turbulence with boundary conditions, for instance in the presence of wall, 
Eqs. (\ref{K0G0}) can not be applied.}

\bigskip

{\bf Remarks}.
Observe that, thanks to the Navier--Stokes bifurcations, the proposed closure formulas modify significantly the mathematical structure of Eqs. (\ref{karman-howarth}) and (\ref{corrsin}).

In Refs. \cite{Khabirov1, Khabirov2}, the authors, studying the non--closed von K\'arm\'an--Howarth equation by group theoretical methods, suggest solutions to the closure problem of isotropic turbulence, especially for what concerns the decay of the turbulence.  
Their two works were not based on the Navier--Stokes bifurcations, and their proposed closure formulas exhibit symmetries.
Here, Eqs. (\ref{K0G0}) may not present such symmetries, and this is due to the Navier--Stokes bifurcations which determine the continuous trajectories divergence and thus the possible absence of these symmetries.

\bigskip

\section{Main properties of the closed equations solutions}

This section analyzes some of the properties of the closed Corrsin and von K\'arm\'an--Howarth equations, with particular reference to the evolution times of the developed kinetic energy and temperature spectra. In particular, we will show that these spectra reach their developed shape in finite times which depend on the the initial condition and on the classical maximal Lyapunov exponent $\Lambda$.

The exponent $\Lambda$ is linked to $\lambda$ through 
\bea
\ds \Lambda \equiv \lim_{r \rightarrow 0} \lambda(r) = \lambda(0)
\label{Lambda}
\eea
and can be calculated with Eq. (\ref{lambda}) considering that, near the origin
$f$ and $f_\theta,$ behave like
\bea
\begin{array}{l@{\hspace{-0.cm}}l}
\ds f(r) = 1 -\frac{1}{2}  \left( \frac{r}{\lambda_T}\right)^2   + ..., \\\\
\ds f_\theta(r) = 1 - \left( \frac{r}{\lambda_\theta} \right)^2 + ...
\end{array}
\label{f_0}
\eea
Hence
\bea
\ds \Lambda(t) = \frac{u(t)}{\lambda_T(t)}
\label{Lambda1}
\eea
Now, with reference to von K\'arm\'an--Howarth and Corrsin equations, $u$ and $\theta$ decrease with $t$ according to \cite{Karman38,  Corrsin_1}
\bea
\begin{array}{l@{\hspace{-0.cm}}l}
\ds \frac{d u^2}{dt} = - \frac{10 \nu u^2}{\lambda_T^2}, \\\\
\ds \frac{d \theta^2}{dt} = - \frac{12 \chi \theta^2}{\lambda_\theta^2}
\label{u theta}
\end{array}
\eea
whereas the correlation scales change following these equations
\bea
\begin{array}{l@{\hspace{-0.cm}}l}
\ds \frac{d \lambda_T}{dt} = 
- \frac{u}{2} + \nu \left( \frac{7}{3} f^{IV}(0) \lambda_T^3 -\frac{5}{\lambda_T}\right), \\\\
\ds \frac{d \lambda_\theta}{dt} = 
- \frac{u}{2} \frac{\lambda_\theta}{\lambda_T} + \chi \left( \frac{5}{6} f_{\theta}^{IV}(0) \lambda_\theta^3 -\frac{6}{\lambda_\theta}\right)
\end{array}
\label{lu ltheta}
\eea
where Eqs. (\ref{u theta}) and (\ref{lu ltheta}) are, respectively, the equations for the coefficients of the powers $r^0$ and $r^2$ of von K\'arm\'an-–Howarth and Corrsin equations.
Equations (\ref{lu ltheta}) include the terms
with $f^{IV}(0)$ and $f_{\theta}^{IV}(0)$ whose determination requires the equations for the coefficients of the power of $r$ higher than 2.
Therefore, we qualitatively discuss the variation laws of the correlation scales and of the classical Lyapunov exponent.
The first terms at the R.H.S. of Eqs. (\ref{lu ltheta}) are responsible for the energy cascade,
whereas the other ones are due to viscosity and thermal diffusivity.
For what concerns the energy cascade, it tends to reduce the correlations scales, and if this is  sufficiently stronger than viscosity and thermal diffusivity effects, 
then $d \lambda_T/dt<0$ and $d \lambda_\theta/dt<0$. On the contrary, the diffusivities tend to increase the correlation scales, thus $f^{IV}(0)$ and $f_{\theta}^{IV}(0)$ are expected to be positive quantities.
\begin{figure}[h]
\centering
\includegraphics[width=0.46000\textwidth]{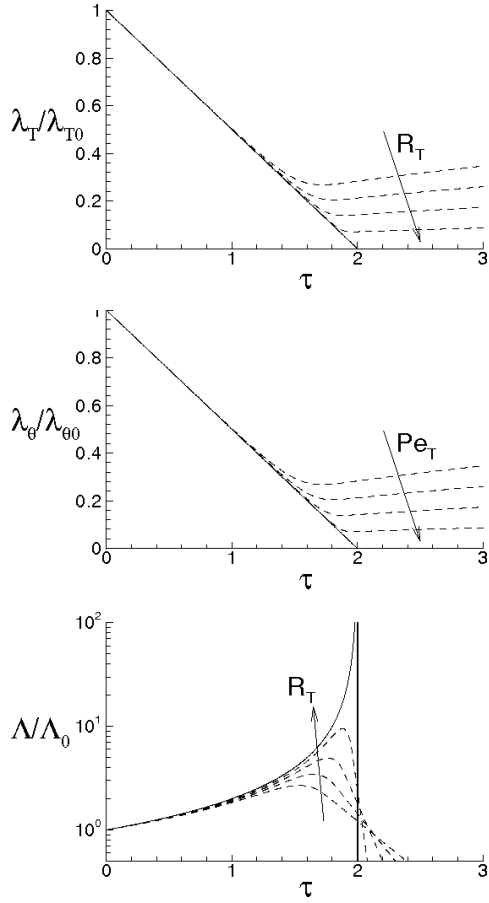}
\caption{Taylor and Corrsin scales, and classical Lyapunov exponent in function of the dimensionless time.}
\label{figura_2}
\end{figure}

For sake of our convenience, the condition $\nu$=0, $\chi$=0 is first analyzed.
In this case, $u$ and $\theta$ are both constants, whereas $\lambda_T$, $\lambda_\theta$ and $\Lambda$
vary with $t$. From Eqs. (\ref{lu ltheta}) and (\ref{Lambda1}), $\lambda_T$ and $\lambda_\theta$ are proportional with each other, and vary linearly with the time according to
\bea
\begin{array}{l@{\hspace{-0.cm}}l}
\ds \frac{\lambda_T(t)}{\lambda_T(0)} \equiv  \frac{\lambda_\theta(t)}{\lambda_\theta(0)} 
 = 1- \frac{\tau}{2}, \\\\
\ds \frac{\Lambda(t)}{\Lambda(0)} = \frac{1}{1-\tau/2}, \\\\
\tau = \Lambda(0) t
\end{array}
\eea
where $\tau$ is the dimensionless time.
Therefore, for $\nu$=0, $\chi$=0, the energy cascade described by 
Eqs. (\ref{K0G0}) determines the decreasing of the Corrsin and Taylor scales until to 
$\tau \rightarrow 2$, where both the spectra are considered to be fully developed, and $\lambda_T \rightarrow 0$, $\lambda_\theta \rightarrow 0$ and  $\Lambda \rightarrow \infty$ 
(see solid lines of Fig. \ref{figura_2}). That is, the correlation functions exhibit developed shapes in
a finite time whose value depends on the initial condition $\Lambda(0)$. The correlations scales are decreasing functions of $\tau$ whereas $u$ and $\theta$ are both constants, and this means that mechanical and thermal energies are transfered from large to small scales.

For $\nu >$0, $\chi>$0, $d u/d\tau <$0 and $d \theta/ d\tau<$0, therefore 
$f$ and $f_\theta$ are considered to be fully developed when $d \lambda_T/dt=0$ and 
$d \lambda_\theta/dt=0$, respectively.
These cases are qualitatively represented by the dashed lines for different values of $R_T$ and $Pe$, where $R_T= \lambda_T u/\nu$, $Pe=Pr R_T$ and  $Pr = \nu/\chi$ are, respectively, the Reynolds number and the P\'eclet number, both referred to the Taylor scale, and the Prandtl number. 
If the initial values of $\lambda_T$ and $\lambda_\theta$ are relatively high, the dissipation effects are quite small in comparison with those of the convective terms. Accordingly, the phenomenon of energy cascade is initially stronger than the diffusivities effects, and this determines that 
$\lambda_T$ and $\lambda_\theta$ diminish exhibiting about the same trend of the case $\nu=\chi$=0.
Following Eqs. (\ref{u theta}) and (\ref{lu ltheta}), the interval of $\tau \in (0, 2)$ can be splitted in two regions for $f$ and $f_\theta$. The first ones occur where $d \lambda_T/d \tau<$0, $d \lambda_\theta/ d\tau<$0 until to  certain values of $\tau_1<$2, $\tau_2<$2, in which 
$d \lambda_T/dt(\tau_1)=0$, $d \lambda_\theta/dt(\tau_2)=0$ (dashed lines) where, in general, $\tau_1 \ne \tau_2$. In this last situation, the mechanism of energy cascade is balanced by viscosity and thermal diffusivity, and the turbulent spectra can be considered to be fully developed. In any case, this happens in finite times 
$\tau <$2 for the two spectra. The Lyapunov exponent initially coincides about with that calculated for $\nu=$0, then reaches its maximum for $\tau_1<2$ and thereafter diminishes due to the viscosity effects.
When $\Lambda$ reaches its maximum, $d\Lambda/ d\tau =$0, it is reasonable that chaos and mixing achieve their maximum level, and the spectra are there considered to be fully developed.

Thereafter, we observe the region where $d \Lambda/ d\tau<$0.
There, due to the smaller values of $\lambda_T$ and $\lambda_\theta$, the dissipation is stronger than the energy cascade, and $\lambda_T$ and $\lambda_\theta$ tend to rise according to 
Eq. (\ref{lu ltheta}). 
This region, which occurs immediately after the fully developed condition ($d\Lambda/ d\tau =$0), corresponds to the decaying turbulence regime.

It is worth to remark that the proposed closure equations (\ref{K0G0}) are expected to be valid 
in the region where $d\Lambda/ d\tau >$0, where the effects of the Navier--Stokes bifurcations generate the regime of  fully developed turbulence.
On the contrary, for decaying turbulence, $d\Lambda/ d\tau <$0, after a certain time, say $\tau^+ \gtrsim \tau_1$, the regime of decaying turbulence may not correspond to the fully developed turbulence, and Eqs. (\ref{K0G0}) are not defined.

\bigskip

{
In the case of complete self--similarity, the several solutions of Eqs. (\ref{karman-howarth})--(\ref{corrsin}) are congruent with each other  by proper scale factors depending on only one of the variables. Thus, the von K\'arm\'an--Howarth and Corrsin equations are  reduced to be ordinary differential equations, and this happens when \cite{Corrsin_2, Karman38}
\bea
\begin{array}{l@{\hspace{-0.cm}}l}
\ds \frac{d \lambda_T^2}{dt}\frac{1}{\nu}=\mbox{const}, \ \ \ \frac{u \lambda_T}{\nu}=R_T = \mbox{const}, \ \ \
\ds \frac{\lambda_\vartheta}{\lambda_T} =\mbox{const}
\end{array}
\label{self_s}
\eea

On the other hand, the proposed closure formulas can not be brought to Eqs. (\ref{self_s}), 
thus Eqs. (\ref{K0G0}) do not give a complete self--preservation. 
Nevertheless, in the cases in which the dimensionless quantities of Eq. (\ref{self_s}) 
exhibit relatively slow variations with respect to the correlations, 
the solutions can be considered to be self--preserved only in first approximation.}
In this case, the correlations read as
\bea
\begin{array}{l@{\hspace{-0.cm}}l}
\ds f = f\left( \frac{r}{\lambda_T(t)}\right), \\\\
\ds f_\theta = f_\theta\left( \frac{r}{\lambda_\theta(t)}\right), \\\\
\ds \frac{\theta(t)}{\theta(0)} = \frac{u(t)}{u(0)}
\end{array}
\eea
Next, from Eq. (\ref{u theta}) we have  
\bea
\ds \frac{\lambda_\theta(t)}{\lambda_T(t)} = \sqrt{\frac{6}{5} \frac{1}{Pr}}
\eea
and $f^{IV}$ is proportional to $f_\theta^{IV}$ according to Eq. (\ref{lu ltheta})
\bea
\ds \frac{f_\theta^{IV}(t, 0)}{f^{IV}(t, 0)} = \frac{7}{3} Pr^2
\eea 
Hence, the correlations are considered to be fully developed when
$\tau^*=\tau_1=\tau_2$. There, $f^{IV}(0,t^*)$ is linked to $\lambda_T(t^*)$ through
Eq. (\ref{lu ltheta}) with $d \lambda_T/dt$=0 \cite{deDivitiis_4}
\bea
\ds f^{IV}(0,t^*) = \frac{3}{7} \frac{1}{\lambda_T^4(t^*)} \left( \frac{R_T(t^*)}{2}+5 \right)  >0
\eea

We conclude this section by observing that the times of developing of $f$ and $f_\theta$ are finite quantities which depend on the initial conditions.

\bigskip

\section{Conclusions}

In order to obtain the closure formulas for von K\'arm\'an--Howarth and Corrsin equations,
this analysis represents the fluid motion in the Lagrangian form. 
As the separation vector varies much more rapidly than the velocity field, 
$\bfxi$ and (${\bf u}, \bftheta$) are assumed to be statistically independent.
This conjecture and the adoption of the Liouville theorem lead to the closure of  von K\'arm\'an--Howarth and Corrsin equations. 
The equations here obtained coincide with those of the previous study \cite{deDivitiis_1, deDivitiis_2}, showing that the approach of Refs. \cite{deDivitiis_1, deDivitiis_2}, dealing with the frame invariance of $K$ and $G$ and the properties of the finite--scale Lyapunov basis, is  in agreement with the present analysis.
This corroborates the previous works, providing a demonstration of the closure formulas completely different with respect to the previous articles.

Thereafter, some of the properties of these equations are studied. In particular, we show that the condition of developed spectra (correlations) in isotropic homogeneous turbulence is reached in finite times whose values depend on the initial conditions.

\bigskip

\section{Acknowledgments}

This work was partially supported by the Italian Ministry for the Universities 
and Scientific and Technological Research (MIUR), and received no specific grant from any funding agency in the public, commercial or not-for-profit sectors.

\bigskip

\end{document}